\documentclass{article}
\usepackage{amsmath}
\usepackage[destlabel=true]{hyperref}
\usepackage{pdfpages}
\usepackage{graphicx}
\usepackage{color}
\usepackage{wrapfig}
\usepackage{cite}
\usepackage{xspace}
\usepackage{pax}
\makeatletter
\PAX@DestReq{7034724.pax}{7}  % Eq. 1
\PAX@DestReq{7034724.pax}{8}  % Eq. 2
\makeatother
\hypersetup{linkbordercolor=green,citecolor=green,}

\newcommand{\eg}{\textit{e.g.}, }
\newcommand{\E}{\mathrm{e}}
\newcommand{\D}{\,\mathrm{d}}
\newcommand{\I}{\mathrm{i}}
\newcommand{\pdc}[3]{\ensuremath{\left(\frac{\partial #1}{\partial #2}\right)_{#3}}}
\newcommand{\ute}{UTe$_2$\xspace}
\newcommand{\dth}{\ensuremath{\boldsymbol{D}_{2h}}\xspace}
\newcommand{\cv}[1]{\ensuremath{\boldsymbol{C}_{#1 v}}\xspace}
\renewcommand{\u}{\ensuremath{\boldsymbol{U}(1)}\xspace}

\date{}

\title{Superconducting Triple Point in \ute: Thermodynamics and Symmetry}

\author{L.A.\,Melnikovsky$\,^{a,b}$\\
$^a\,$Weizmann Institute of Science, Rehovot, Israel\\
$^b\,$P.L.\,Kapitza Institute for Physical Problems, RAS, Moscow, Russia
}

\begin{document}
\maketitle

\begin{abstract}
Three lines of second-order phase transitions between the normal phase and two distinct superconducting phases meet at a single point on the phase diagram of \ute \cite{BVK}. Although it is commonly believed otherwise, such triple points are not subject to thermodynamic constraints. In \cite{MM} the phase diagram is interpreted within Landau theory in terms of two superconducting order parameters with different gauge symmetries. Such an interpretation is unique under the assumption of spatial uniformity.
\end{abstract}

\section{Introduction}
\begin{wrapfigure}[21]{r}{45mm}
\begin{center}
  \includegraphics{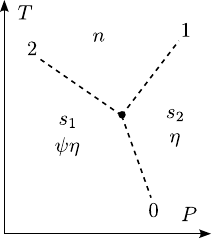}
\end{center}
\caption{Schematic phase diagram with a triple point on the second-order transition lines separating the normal phase $n$ and two superconducting phases $s_1$ and $s_2$ also denoted by $\psi\eta$ and $\eta$.}
\label{fig}
\end{wrapfigure}
Uranium ditelluride is an unconventional superconductor with a rich phase diagram. According to calorimetric measurements~\cite{BVK}, the normal phase and two distinct superconducting phases are separated by three lines of second-order phase transitions that meet at a single point. Such a triple point is widely regarded as subject to strong thermodynamic constraints. Significant efforts have been made to reconcile experimental observations with this belief by questioning the continuous character of the transitions, by matching the slopes of the transition lines, or by introducing a hidden transition line~\cite{BVK,Thomas2021,Erratum2021,Rosuel2023,Sakai2023,Tokiwa2025,Vasina2025,Wu2025,Kamat2026,Vasina2026,Valiska2026}. While we cannot rule out any of these alternatives, Sec.~\ref{therm} shows that they are not inevitable.

A constructive scenario based on Landau theory, which may explain the triple point of the type observed in \ute, was proposed by Marchenko and the present author~\cite{MM}. Two different superconducting order parameters, $\psi$ and $\eta$, were assumed to transform differently under gauge transformations: if $\psi$ is multiplied by $\E^{2\I k\phi}$, then $\eta$ is multiplied by $\E^{2\I mk\phi}$, with $m>1$. In Sec.\ref{sym} we show by symmetry analysis that this identification is in fact unique for spatially uniform system.

\section{Thermodynamic considerations}
\label{therm}
Thermodynamic constraints on the phase diagram with a triple point on the second-order transition lines are usually derived from the Ehrenfest relations~\cite{LL5}, which connect the jumps of the thermodynamic derivatives
\begin{equation}
\alpha = \pdc{S}{T}{P},\quad \beta = \pdc{V}{T}{P}, \quad \gamma = -\pdc{V}{P}{T}
\end{equation} 
on a transition line to its slope $p=\D P/\D T$:
\begin{equation}
\Delta\beta = p \Delta\gamma, \qquad\qquad\quad
\Delta\alpha = p^2 \Delta\gamma.
\end{equation}
Near the triple point Fig.~\ref{fig}, three transition lines meet. We use the subscripts 0, 1, and 2 to refer to the transition lines $s_1|s_2$, $n|s_2$, and $n|s_1$, respectively, while the superscripts are reserved exclusively for the powers:
\begin{equation}
p_0=\frac{\D P_0}{\D T}, \qquad
p_1=\frac{\D P_1}{\D T}, \qquad
p_2=\frac{\D P_2}{\D T},
\end{equation} 
\begin{equation}
\begin{aligned}
\Delta_0\beta &= p_0 \Delta_0\gamma,
&\qquad\qquad\quad
\Delta_0\alpha &= p_0^2 \Delta_0\gamma,\\
\Delta_1\beta &= p_1 \Delta_1\gamma,
&
\Delta_1\alpha &= p_1^2 \Delta_1\gamma,\\
\Delta_2\beta &= p_2 \Delta_2\gamma,
&
\Delta_2\alpha &= p_2^2 \Delta_2\gamma.
\end{aligned}
\end{equation}

If the sums of the corresponding jumps on the three lines
\begin{equation}
\tag{*}
\begin{alignedat}{8}
&\Delta_0\gamma+&&\Delta_1\gamma+&&\Delta_2\gamma&&=
 p_0^0 \Delta_0\gamma &&+ 
 p_1^0 &&\Delta_1\gamma &&+
 p_2^0 &&\Delta_2\gamma,\\
&\Delta_0\beta+&&\Delta_1\beta+&&\Delta_2\beta&&=
  p_0^1\Delta_0\gamma &&+
  p_1^1&&\Delta_1\gamma &&+
  p_2^1&&\Delta_2\gamma
,\\
&\Delta_0\alpha+&&\Delta_1\alpha+&&\Delta_2\alpha&&=
  p_0^2\Delta_0\gamma &&+
  p_1^2&&\Delta_1\gamma &&+
  p_2^2&&\Delta_2\gamma,
\end{alignedat}
\end{equation}
were to vanish, as they would at a first-order triple point, one would obtain a homogeneous linear system in $\Delta_0\gamma$, $\Delta_1\gamma$, and $\Delta_2\gamma$, whose consistency requires the Vandermonde determinant to vanish; that is, two of the slopes $p_0$, $p_1$, and $p_2$ would coincide. In that case, the jumps on the corresponding two transition lines would be equal in magnitude, while the jump on the third line would vanish.

The key point is that second-order phase transitions are singularities of thermodynamic quantities rather than points of phase coexistence~\cite{LL5}. Therefore, the limiting value of a thermodynamic quantity may depend on the path by which the triple point is approached within one phase. For example, the same symbol $\gamma_{\psi\eta}$ appearing in the expressions
\begin{equation}
\tag{**}
\Delta_2\gamma = \gamma_{n}-\gamma_{\psi\eta}, \qquad 
\Delta_0\gamma = \gamma_{\psi\eta} - \gamma_{\eta},
\end{equation}
represents two \emph{distinct} limits of $\gamma_{\psi\eta}(P,T)$, taken along lines 2 and 0, respectively. As a result, apparent zero-sum identities for the jumps of thermodynamic quantities may not hold:
\vspace{-3mm}
\begin{equation}
\begin{alignedat}{4}
&\Delta_0\gamma+&&\Delta_1\gamma+&&\Delta_2\gamma &&\ne 0,\\
&\Delta_0\beta+&&\Delta_1\beta+&&\Delta_2\beta &&\ne 0,\\
&\Delta_0\alpha+&&\Delta_1\alpha+&&\Delta_2\alpha &&\ne 0.
\end{alignedat}
\end{equation}
Therefore, triple points on second-order transition lines are not constrained by thermodynamics.

This path dependence of limiting thermodynamic values is already present within Landau theory and is unrelated to divergent fluctuations. In particular, for the Ginzburg-Landau expansion assumed in Ref.~\cite{MM} [Eqs.~(\hyperlink{PAX@7034724.pax@7}{1}) and (\hyperlink{PAX@7034724.pax@8}{2}) of that work],
\begin{equation}
(at + vp)|\psi|^2 + A|\psi|^4 + (bt - wp)|\eta|^2 + B|\eta|^4 + C|\psi|^2 |\eta|^2 + I(\psi^{*m} \eta + \psi^m \eta^*),
\end{equation}
where $m\ge 4$, the non-analytic part of $\alpha$ (which determines the specific heat) in the $\psi\eta$ phase has the following {\em finite} limits along the transition lines $\psi\eta|n$ and $\psi\eta|\eta$:
\vspace{-1mm}
\begin{equation}
  \alpha_{\psi\eta} \xrightarrow[\ \psi\eta|n\ ]{} \frac{a^2}{2A}, \qquad\qquad
  \alpha_{\psi\eta} \xrightarrow[\ \psi\eta|\eta\ ]{} 2\frac{Ab^2+Ba^2-abC}{4AB-C^2}.
\end{equation}

\vspace{-3mm}
\section{Symmetry considerations}
\label{sym}
Second-order transitions separate phases of different symmetry. The symmetry transformations of the normal non-magnetic phase are compositions of optional time reversal $R$, spatial transformations, and gauge transformations. Gauge transformations form a continuous group \u of complex numbers with unit modulus, which acts on the superconducting order parameter by multiplication. Unless one of the transitions is associated with translational-symmetry breaking (\eg by the formation of a charge-density wave), it is sufficient to take the crystal class of uranium ditelluride, \dth, as the spatial symmetry group of the normal phase. The relevant symmetry group is a direct product:
\begin{equation}
{\cal G}=\dth\times \{\u \cup R \u\}.
\end{equation}
Symmetries of the superconducting phases $s_1$ and $s_2$ are lower, their 
symmetry groups ${\cal G}_1$ and ${\cal G}_2$ form a chain of subgroups
\begin{equation}
\label{chain}
  {\cal G}_1 < {\cal G}_2 < {\cal G}. 
\end{equation}
This follows~\cite{LL5} from the experimental observation~\cite{BVK} that the specific heat decreases across the transitions $s_1 \rightarrow s_2 \rightarrow n$.

Symmetry breaking at a second-order phase transition is described by an order parameter that transforms according to an irreducible representation of the higher-symmetry group. Order parameters $\psi$ and $\eta$ of the superconducting phases $s_1$ and $s_2$ transform according to different irreducible representations of the group ${\cal G}$.

\dth is an elementary abelian 2-group\footnote{A finite commutative group in which every non-identity element has order 2.} with one-dimensional irreducible representations. The group $\{\u \cup R \u\}$ is isomorphic to \cv{\infty} (under the identification of \u with rotations and $R$ with reflection). All of its non-unit physically irreducible representations are two-dimensional and are labeled by natural numbers~\cite{LL3}. Suppose that $\psi$ transforms according to the irreducible representation of \cv{\infty} labeled by $2k$ (corresponding to composite bosons built from $2k$ electrons) and according to a representation $\chi$ of \dth:
\begin{itemize}
  \item
    If $\chi(g)\equiv 1$ (that is $\chi$ is the unit representation), then the symmetry group of phase $s_1$ is
    \begin{equation}
      \label{eq:G-unit}
      {\cal G}_1 = \cv{2k} \times \dth,
    \end{equation}
    $\psi$ does not change under spatial transformations in \dth and the spatial symmetry remains the same as in the normal phase;

  \item
    If $\chi(g) \ne 1$ for some $g \in \dth$, then the spatial symmetry is reduced to the subgroup (of index 2) $\ker(\chi) < \dth$ and
    \begin{equation}
      \label{eq:G-entangled}
      {\cal G}_1 = \left(\cv{2k} \times \ker(\chi)\right) \ \cup\ C_{4k} g \left(\cv{2k} \times \ker(\chi)\right).
    \end{equation}
\end{itemize}
The same analysis applies to the order parameter $\eta$ and to the symmetry group~${\cal G}_2$.

Eqs.~\eqref{eq:G-unit} and \eqref{eq:G-entangled} imply that the subgroup chain~\eqref{chain} is possible only if $\eta$ transforms according to the representation of \cv{\infty} labeled by $2mk$, with $m>1$. In other words, as suggested in~\cite{MM}, if $\psi$ corresponds to a $2k$-electron condensate, then $\eta$ corresponds to a $2mk$-electron condensate. For odd $m$, $\eta$ transforms according to the same representation $\chi$ as $\psi$, and phases $s_1$ and $s_2$ have the same spatial symmetry. For even $m$, $\eta$ is invariant under \dth, and phases $n$ and $s_2$ have the same spatial symmetry.

I thank V.I.\,Marchenko, S.S.\,Sosin, and E.V.\,Surovtsev for useful discussions. I am also grateful to the anonymous referees, whose feedback prompted more extensive exposition and detailed presentation.

\includepdf[pages=-,
	    noautoscale=true,
      pagecommand*={\vspace*{-8.4mm}\section*{Appendix}\label{section*.2}\vspace*{-11mm}},
      pagecommand={
        \noindent\setlength{\unitlength}{1mm}
        \begin{picture}(0,0)\put(52,-208){\color{white}\rule{15mm}{15mm}}\end{picture}
        },
      ]{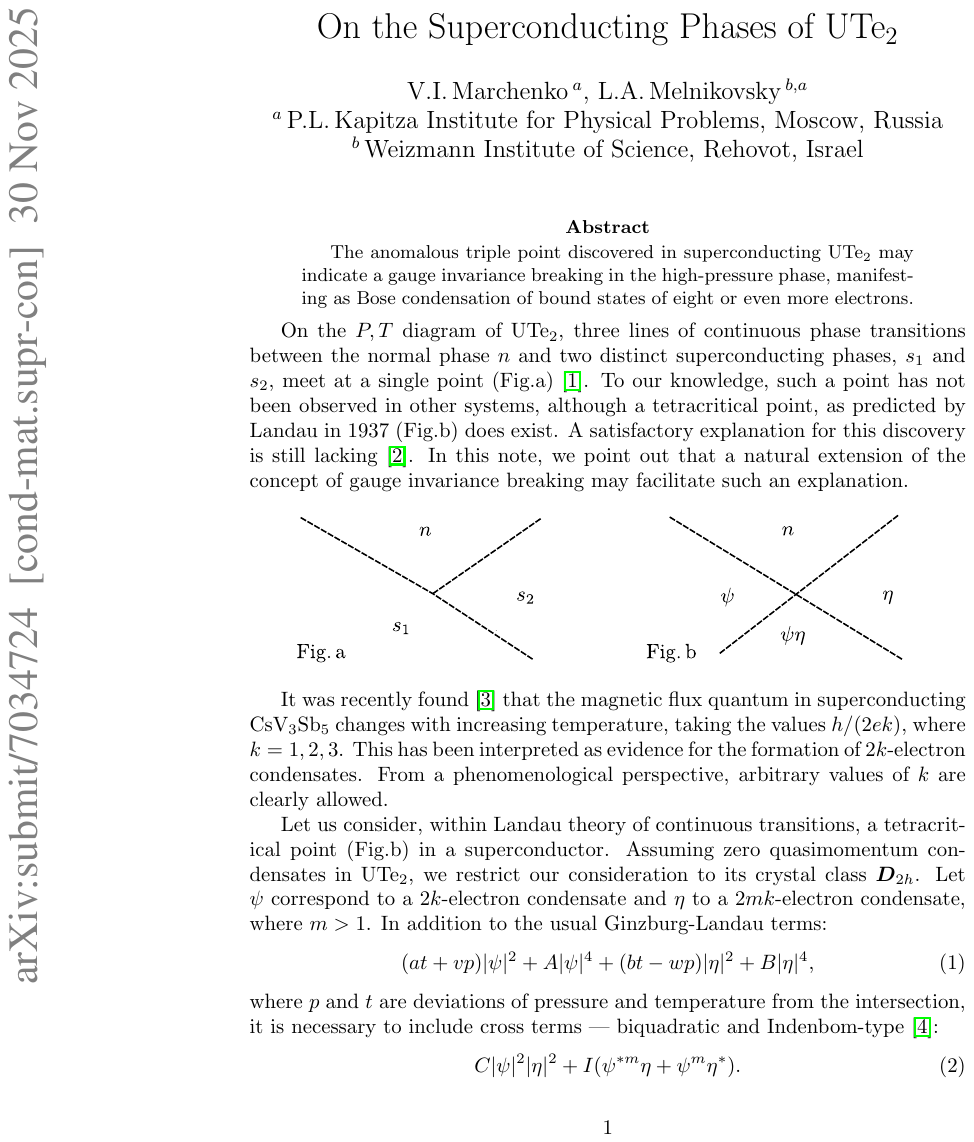}

\end{document}